%% file: main.tex
\documentclass[sigconf,authorversion]{acmart}
\usepackage{amsmath,amsfonts}
\usepackage{algorithmic}
\usepackage{graphicx}
\usepackage{textcomp}
\usepackage{xcolor}
\usepackage{xspace}
\usepackage{hyperref}
\usepackage{tabularray}
\usepackage{verbatim}
\usepackage[frozencache,cachedir=.]{minted}
\usepackage{booktabs}
\usepackage{amsmath}
\usepackage{lipsum}
\usepackage{array}
\usepackage[acronym,nohypertypes={acronym}]{glossaries}
\usepackage{orcidlink}

\definecolor{LightGray}{gray}{0.9}
\setminted{breaklines}
\setminted{bgcolor=LightGray}

\newcommand{\mm}[1]{{\small{\textsf{#1}}}}
\newcommand{\etal}[0]{\emph{et al.}\xspace}
\newcommand\ie{\emph{i.e.},\xspace}
\newcommand\eg{\emph{e.g.},\xspace}

\newcommand{\citesec}[1]{Section~\ref{sec:#1}}

\newacronym{cdf}{CDF}{\textit{Cumulative Distributed Functions}}
\newacronym{llm}{LLM}{\textit{Large Language Model}}

\usepackage{changepage}
\usepackage{framed}
\makeatletter
 {\par\unskip\endMakeFramed}
\makeatother

\definecolor{formalshade}{rgb}{0.93,0.93,0.93}
\definecolor{darkblue}{rgb}{0.2, 0.2, 0.2}

\newenvironment{formal}{%
  \def\FrameCommand{%
    \hspace{1pt}%
    {\color{darkblue}\vrule width 2pt}%
    {\color{formalshade}\vrule width 4pt}%
    \colorbox{formalshade}%
  }%
  \MakeFramed{\advance\hsize-\width\FrameRestore}%
  \noindent\hspace{-1pt}%
  \begin{adjustwidth}{}{7pt}%
  \vspace{2pt}\vspace{2pt}%
}
{%
  \vspace{3pt}\end{adjustwidth}\endMakeFramed%
}

\newcounter{resultcounter}

\copyrightyear{2024}
\acmYear{2024}
\acmBooktitle{28th International Conference on Evaluation and Assessment in Software Engineering (EASE 2024), June 18--21, 2024, Salerno, Italy}
\acmDOI{10.1145/3661167.3661221}
\acmISBN{979-8-4007-1701-7/24/06}

\setcopyright{none}

\begin{document}
\title{
A Performance Study of \acrshort{llm}-Generated Code on Leetcode}

\author{Tristan Coignion}
\affiliation{%
    \institution{Univ.\,Lille, CNRS, Inria}
    \country{France}
}
 \orcid{0009-0003-2525-3637}
\email{tristan.coignion@inria.fr}

\author{Cl\'ement Quinton}
\affiliation{%
    \institution{Univ.\,Lille, CNRS, Inria}
    \country{France}
}
\orcid{0000-0003-3203-6107}
\email{clement.quinton@inria.fr}

\author{Romain Rouvoy}
\affiliation{%
    \institution{Univ.\,Lille, CNRS, Inria}
    \country{France}
}
\orcid{0000-0003-1771-8791}
\email{romain.rouvoy@inria.fr}

\begin{abstract}
This study evaluates the efficiency of code generation by \glspl{llm} and measures their performance against human-crafted solutions using a dataset from Leetcode.
We compare 18 \glspl{llm}, considering factors such as model temperature and success rate, and their impact on code performance. 
This research introduces a novel method for measuring and comparing the speed of \gls{llm}-generated code, revealing that \glspl{llm} produce code with comparable performance, irrespective of the adopted \gls{llm}.
We also find that \glspl{llm} are capable of generating code that is, on average, more efficient than the code written by humans. 
The paper further discusses the use of Leetcode as a benchmarking dataset, the limitations imposed by potential data contamination, and the platform's measurement reliability. 
We believe that our findings contribute to a better understanding of \gls{llm} capabilities in code generation and set the stage for future optimizations in the field.
\end{abstract}

\maketitle

\glsresetall
\input{Introduction}
\input{Methodology}
\input{Results}
\input{Discussion}
\input{Related}
\input{Conclusion}

\begin{acks}
This work received support from the French government through the {\em Agence Nationale de la Recherche} (ANR) under the France\,2030 program, including partial funding from the {\sc CARECloud}~({\sf ANR-23-PECL-0003}), DISTILLER~({\sf ANR-21-CE25-0022}), and KOALA~({\sf ANR-19-CE25-0003-01}) projects.
Experiments presented in this paper were carried out using the Grid'5000 testbed, supported by a scientific interest group hosted by Inria and including CNRS, RENATER and several Universities as well as other organizations.\footnote{See \url{https://www.grid5000.fr}}
\end{acks}

\bibliographystyle{ACM-Reference-Format}
\bibliography{ase}

\end{document}

%% file: Introduction.tex
\section{Introduction}
\glspl{llm} have recently increased in popularity, especially with the advent of ChatGPT~\cite{openaiGPT4TechnicalReport2023}.
While \glspl{llm} have been spreading over various application domains, such as text or image generation, certain types of \glspl{llm} are being developed solely for code-related purposes.
These \glspl{llm} aim to assist the developers by saving time and effort through the generation of code, documentation, unit tests, etc.
Many of these \glspl{llm} only come in a ``raw" form, that is, they do not integrate themselves into the developer's coding process.
These include models, such as {\sc CodeGen}~\cite{nijkampCodeGenOpenLarge2022}, %
{\sc StarCoder}~\cite{li2023starcoder}, 
{\sc WizardCoder}~\cite{luoWizardCoderEmpoweringCode2023}, 
{\sc CodeT5}~\cite{wangCodeT5IdentifierawareUnified2021}, and {\sc Incoder}~\cite{friedInCoderGenerativeModel2022}.
On the other hand, some \glspl{llm} are already seamlessly integrated into the developer's IDE as code assistants, like {\sc GitHub Copilot},\footnote{\url{https://github.com/features/copilot}} {\sc Amazon CodeWhisperer},\footnote{\url{https://aws.amazon.com/fr/codewhisperer/}} and {\sc Tabnine}.\footnote{\url{https://www.tabnine.com/}}

There has been a significant amount of work dedicated to comprehending how these \glspl{llm} perform in various situations and defining their limits. 
For instance, several works address the security of the code generated by such models~\cite{pearceAsleepKeyboardAssessing2022, sandovalSecurityImplicationsLarge2022, perryUsersWriteMore2022} or the prevalence of bugs in the generations~\cite{jesseLargeLanguageModels2023}. 
Many researchers are also investigating how developers interact with \glspl{llm} and how such models fit into the programming workflow~\cite{perryUsersWriteMore2022, vaithilingamExpectationVsExperience2022, barkeGroundedCopilotHow2022} 
There is also a broad research effort to measure the actual efficiency of these \glspl{llm} by creating common benchmarks for comparison~\cite{chenEvaluatingLargeLanguage2021, austinProgramSynthesisLarge2021, yuCoderEvalBenchmarkPragmatic2023, hendrycksMeasuringCodingChallenge2021, wangReCodeRobustnessEvaluation2022} or by actually measuring different qualities related to the generations, such as the success rate~\cite{yetistirenAssessingQualityGitHub2022} or the robustness of the model regarding variations~\cite{doderleinPilotingCopilotCodex2022}.

To the best of our knowledge, there is no research work evaluating the performance of the code generated by \glspl{llm}.
Yet, having code that runs faster is an often sought-after characteristic of a program. 
Indeed, programming efficiency is paramount, especially when resources are scarce or programs are deployed on a large scale. 
In today's context where the energy consumption of software systems has become a major concern,
improving software efficiency is particularly relevant since increasing the performance of a program can also lead to energy consumption reduction~\cite{verdecchiaEstimatingEnergyImpact2017, acar:hal-01496266}.

The process of code optimization is lengthy and intricate, requiring careful attention and a certain level of expertise, especially when searching for the best-performing algorithm, selecting the most appropriate data structure, or struggling with memory hierarchy.
Yet, this process is necessary to identify opportunities for improvement that may result in minor reductions in execution time.
\glspl{llm} can be used as a way to make this process easier, \eg by generating performance-improving code edits~\cite{madaanLearningPerformanceImprovingCode2023, gargDeepDevPERFDeepLearningbased2022, chenSupersonicLearningGenerate2023}. 
Garg~\etal~\cite{gargRAPGenApproachFixing2023} also presented {\sc RAPGen}, a method for generating zero-shot prompts to enhance performance.
However, while users seem to put a lot of trust in code generated by \glspl{llm}, they still have trouble reviewing it~\cite{vaithilingamExpectationVsExperience2022, sandovalSecurityImplicationsLarge2022}, which can lead to slow code being shipped in production, especially if an \gls{llm} generates inefficient code.

The key contributions of this paper are as follow :
(i) We study the performance of the code generated by 18 \glspl{llm} on 204 problems and investigate performance differences across models using a novel method for measuring and comparing the performance of \gls{llm}-generated code.
(ii) We compare the performance of the code generated by \glspl{llm} to the code written by humans.
(iii) Incidentally, we evaluate the usability of Leetcode,\footnote{\url{https://leetcode.com/}} a public repository of algorithmic problems that we use as a dataset.

From \autoref{sec:method} to \autoref{sec:data_analysis}, we describe the tasks dataset and the models we selected, outline our experiment setup, and explain the methodology we followed to analyze the obtained results, respectively.
We report in \autoref{sec:results} on the results of our evaluation and provide a critical discussion in \autoref{sec:discussion}. 
Finally, \autoref{sec:rw} presents the related works, and \autoref{sec:conclusion} concludes the paper.

%% file: Methodology.tex
\section{Methodology}
\label{sec:method}

\subsection{Research questions}
This paper covers the performance of code generated by various \glspl{llm}.
In particular, we aim to answer the following research questions:

\textbf{RQ1}: {\em Can Leetcode be used as a dataset and a benchmark platform for evaluating \glspl{llm}?}
Leetcode can serve as both a dataset of problems and a tool to evaluate and measure solutions to the problems.
Particularly, we study if the dataset is subject to recitation and if the measures Leetcode provides are reliable.

\textbf{RQ2}: {\em Are there notable differences between the performance of the code generated by different \glspl{llm}?}
\glspl{llm} differ greatly in terms of generating correct code, so we want to know if they also differ in terms of generating efficient code.

\textbf{RQ3}: {\em Is there an effect of the success rate and the temperature of the \gls{llm} on the code's performance?}
Having a higher temperature decreases the capacity of the \glspl{llm} to generate valid code, so we aim to study if this also applies to the performance of the code. 
In the same way, we also study if an \gls{llm} that is very good at generating valid code on one problem, is going to generate efficient solutions to this problem.

\textbf{RQ4}: {\em How efficient are the solutions generated by the \glspl{llm} compared to human solutions?}
Comparing the \glspl{llm} to a set of human-authored solutions can provide insights into their position relative to humans in terms of code performance.

\subsection{Tasks \& Dataset}\label{sec:tasks}

\textbf{Task selection}.
The input tasks---\emph{i.e.}, problems specified by a prompt---we consider to generate code from various \glspl{llm} has to meet the following requirements:
\begin{itemize}
    \item A given problem, should offer multiple candidate solutions, whose generated code performs differently.
    This ensures one can observe differences across the various \glspl{llm};
    \item Generated solutions should exhibit variable execution times. 
    Given more complex inputs, one should differentiate $\mathcal{O}(n)$ from $\mathcal{O}(2n)$ and $\mathcal{O}(n^2)$ algorithms.
\end{itemize}
As a result, task datasets, such as {\sc HumanEval} or {\em Mostly Basic Python Programming} (MBPP), which are classically used when evaluating \glspl{llm} for code assessments~\cite{chenEvaluatingLargeLanguage2021, nijkampCodeGenOpenLarge2022, CodeparrotCodeparrotHugging2022, allalSantaCoderDonReach2023}, cannot be considered for our purpose.
Indeed, while they do provide unit tests to drive the generations, the size of the inputs in these unit tests remains small and fails to scale to appreciate performance issues.
Moreover, the solutions that need to be generated are often very short, which would lead to fewer possible variations between implementations.
Also, the fact that many problems are not algorithmic by nature makes them less prone to inefficient practices and performance variation.
To face such issues, we used input prompts that were built from Leetcode problems.
Leetcode is an online judge platform that suggests programming problems to registered users.
It addresses the above limitations as it provides algorithmic problems with varying levels of difficulty and test cases with large input sizes.
Leetcode also exposes a {\sc GraphQL\,API}\footnote{\url{https://pypi.org/project/python-leetcode/}} to fetch relevant metadata on the problems, such as exercise instructions, code snippets containing the signature of the function to generate, as well as the difficulty and topics of the problem.

We followed an experimental design similar to the one used by Döderlein~\etal~\cite{doderleinPilotingCopilotCodex2022}, while using a different set of Leetcode questions. 
To avoid data contamination, which happens when an \gls{llm} is tested on data it was trained on,\footnote{You can find more details on data contamination \href{https://web.archive.org/web/20231020160612/https://bdtechtalks.com/2023/07/17/llm-data-contamination/}{here}.} we only considered problems that were published after January $1^{st}$, 2023. 
As all the \glspl{llm} (except {\sc GitHub Copilot}) we evaluated were trained using datasets older than these problems, we avoid any data contamination.
As these problems are published by Leetcode in the context of programming competitions, they are always original.
However, {\sc GitHub Copilot} being an online closed-source tool, one cannot tell whether it underwent training with the problem set we employed. 
Our set was composed of 204 problems---labeled as 56 easy problems, 104 medium problems, and 44 hard problems, as classified by Leetcode upon the problem's publication.

To answer RQ1, we also performed our experiment a second time on the set of questions used by Döderlein~\etal~\cite{doderleinPilotingCopilotCodex2022}, which is composed of 300 problems (95 easy, 105 medium, and 100 hard) from the most liked problems of Leetcode. 
We will refer to this dataset as the "old" dataset, and to our dataset of problems published during 2023 as the "new" dataset.
Code generation was performed in Python
for its ease of use and because of the prevalence of Python-written datasets for evaluating \glspl{llm}~\cite{chenEvaluatingLargeLanguage2021, austinProgramSynthesisLarge2021}.

\begin{figure}[b!]
    \centering
    \begin{minted}[
        escapeinside=||,
        framesep=2mm,
        fontsize=\footnotesize,
        ]{python}
# Start of the input prompt
"""
Given an integer array nums, return all the triplets `[nums[i], nums[j],nums[k]]` such that `i != j`, `i != k`, and `j != k`, and `nums[i] + nums[j] + nums[k] == 0`.   

Notice that the solution set must not contain duplicate triplets.
"""
class Solution:
    def threeSum(self, nums: List[int]) -> List[List[int]]:
        # End of the input prompt 
        
        # Start of the generated code
        nums.sort()
        res = []
        for i in range(len(nums) - 2):
            ...
        # End of the generated code

# Start of the benchmarking code
def check():
    Solution().threeSum([82597, -9243, 83030, ...])
    Solution().threeSum([0, 0, 0, 0, ...])
    Solution().threeSum([0, 0, -1, -1, ...])

import pytest
@pytest.mark.benchmark(group="3sum")
def test_3sum_generated_1(benchmark):
    benchmark(check)
# End of the benchmarking code
\end{minted}
    \caption{Example of problem's input prompt, generated code, and benchmarking code.}
    \label{fig:code_example}
\end{figure}

\textbf{Input prompts}.
The instructions given by Leetcode for each programming problem contain \emph{(i)}~the description of the problem, \emph{(ii)}~examples of inputs and outputs, and \emph{(iii)}~constraints on the input data. 
To build the prompts inputted to the \glspl{llm}, we chose to only include the description of the problem.
This choice aimed to maintain the prompt's conciseness and ensure compliance with potential \glspl{llm}' length restrictions. %
Indeed, \glspl{llm} are constrained by a context size, imposing the maximum number of tokens they can process at once, encompassing both the prompt and the answer. 
In our study, the majority of \glspl{llm} impose a context window of 1024 tokens.
\autoref{fig:code_example} illustrates a descriptive prompt (enclosed within triple quotes) along with a solution generated using this prompt and the corresponding benchmark instructions. 
Leetcode occasionally includes an additional comment in the prompt, indicating the available methods of the elements passed to the solution (\eg binary trees).
While adding examples in the prompt could potentially increase the likelihood of generating correct solutions~\cite{doderleinPilotingCopilotCodex2022}, we chose not to include them due to the complexity of representing Leetcode's data structures (such as arrays, graphs, linked lists, trees, etc.) in textual form.
On Leetcode's website, data structures are textually represented using an array notation format with brackets (\eg ``{\tt [1, 2, 3]}"), which might be misleading for an \gls{llm} working with data structures that differ from arrays or lists.

\textbf{Canonical solutions}.
Each problem was matched with a single valid solution, written by a human and fetched from various sources.
These solutions, referred to as ``canonical solutions", are considered as baselines during the benchmarking process, although they may not represent the entirety of human-written solutions. 
They were used to assess the stability of the measuring process.
Most of the canonical solutions were fetched from the WalkCC repository of Leetcode solutions.\footnote{\url{https://github.com/walkccc/LeetCode}}
When one question did not come up with any solution in Python from this repository, we selected one solution proposed by the Leetcode community among the most upvoted ones (starting from the one with the most upvotes), which are also publicly available on the Leetcode website.
These modifications only included changes to the function name, variable names, and type hints except for one specific case, where a recursive solution provided by WalkCC was replaced by an iterative one from the Leetcode community because of the stack limit on our local setup.
We ended up with one Python-based canonical solution for each problem of our dataset.

\textbf{Test cases}.
Testing generated solutions is a twofold process, as both the correctness and scalability (performance) of each solution must be checked.
First, to ensure that generated solutions are correct, we provide such solutions to the Leetcode online judge system, which in turn validates them based on its test suites.
Second, we execute the valid solutions with input data to assess their performance.
To retrieve such input data, we crawled through the problems' instructions and extracted two to three examples of inputs and expected outputs provided by Leetcode.
However, such inputs were too small to exhibit significant performance differences, \eg arrays containing only two to five elements.
To execute the generated solutions with larger input data, we took advantage of Leetcode's judge system that returns the inputs and expected output of the first failed test of a test suite.
We noticed that Leetcode tests if the submitted solution is inefficient by using a timeout system. 
Since all selected problems are algorithmic by nature, one simple way to put a heavier load on their implementations is to increase the size of the inputs.
For every problem, we thus submitted a modified version of a canonical solution that failed only when the size of the first parameter exceeded a certain threshold. 
We then set this threshold manually multiple times for every problem to extract three different inputs for each problem, resulting in more than 150\,MB of fetched input data with most inputs having over $10^5$ elements.

\subsection{\glspl{llm} Under Study}
Our empirical study covers a total of $18$ \glspl{llm}, specifically designed for coding purposes.
We selected the $18$ popular code \glspl{llm} from Hugging Face,\footnote{\url{https://huggingface.co}} as well as {\sc GitHub Copilot}, which is an online closed-source code assistant. The LLMs were selected based on the number of downloads and likes they exhibited.
We chose {\sc GitHub Copilot} to offer a comparison between a commercial \gls{llm} and open-source \glspl{llm}, but did not choose any other GPT models from OpenAI because of their cost.
\autoref{table:models} summarizes all the \glspl{llm} considered in this study.
This includes variants of the same base \gls{llm}---\ie models with varying sizes (in billions of parameters) or different training data, such as variants of {\sc CodeGen}, {\sc InCoder}, and {\sc CodeT5}.
\glspl{llm} belonging to the same family are models closely related in terms of training data and method. 
We first performed our experiment in March 2023 with a subset of the models presented here, with the "old" dataset. 
We then performed it a second time in September~2023 with all the models and the "new" dataset.

\begin{table}[ht] \centering
    \renewcommand{\arraystretch}{0.63}
    \begin{small}
        \input{models.tex}

    \end{small}
    \vspace{2mm}
    \caption{\glspl{llm} considered in our study. Models with the RQ1 checkmark were also evaluated on the "old" dataset. Size is in billions of parameters}
    \label{table:models}
\end{table}
\vspace{-10mm}

\section{Experiment Setup}
This section describes our experiment setup to generate and validate the solutions produced by the \glspl{llm}. 
First, we describe how solutions were generated from each \gls{llm}.
Then, we outline the three-step process used to filter invalid solutions. 
Finally, we explain how the run time of the generated solutions was measured.

\subsection{Code Generation}
We generated 10 solutions for each problem from our dataset by varying the temperature of the \glspl{llm} used to generate them. 
Specifically, we considered 6 different temperatures ({\tt 0.1}, {\tt 0.2}, {\tt 0.4}, {\tt 0.6}, {\tt 0.8}, and {\tt 1.0}).
As {\sc Copilot}'s temperature cannot be configured, we used its default temperature for generating all the problems.

\textbf{Generating code with {\sc GitHub Copilot}}. 
Automatically generating with {\sc GitHub Copilot} for a reproducible experiment proved to be a difficult task.
Firstly, the code suggestion feature of {\sc Copilot} activates when typing in a text editor with the installed {\sc Copilot} plugin. 
Additionally, {\sc GitHub Copilot} produces code based on a context that encompasses the current file and the files previously accessed by the user, impacting the generated solutions.
Lastly, we noticed a caching mechanism on the {\sc GitHub Copilot} server side, which resulted in very similar or identical solutions if we generated multiple solutions for a given problem in the same session.
To address these issues, we used a generation method similar to the one used by Döderlein~\etal~\cite{doderleinPilotingCopilotCodex2022} by instrumenting the {\sc GitHub Copilot} Neovim plugin\footnote{https://github.com/github/copilot.vim} and restarting the plugin between every generation to avoid the caching effect.
This method allowed us to automatically generate solutions in a quick and isolated fashion.
On top of that, {\sc GitHub Copilot} provides 2 means of generation: {\em inline} generations (the suggested code is integrated with the editor) and {\em panel} generations ({\sc Copilot} generates at most 10 completions and displays them on a panel next to the editor).
We chose to exclusively use inline generations, as panel generations yielded worse results in terms of functional correctness than inline generations.

\textbf{Generating code with open-source models}.
Regarding the open-source models, we generated solutions by deploying the models on servers provided by the {\sc Grid5000} platform~\cite{grid5000}.
We used the {\sf Deepspeed} library to make the generation process faster and fit larger models on our GPUs.
Concerning the sampling, we used the same methods as Chen \etal~\cite{chenEvaluatingLargeLanguage2021} and used nucleus-sampling~\cite{holtzmanCuriousCaseNeural2020} with top $p=0.95$. 
The maximum number of tokens to be generated was set to $600$. 
This is because the \glspl{llm} we used have a limited context size of 1024 tokens (prompt included). 
Thus, to avoid exceeding the context size, we had to limit the number of tokens to generate.
We also verified it did not significantly impact the functional validity of the \glspl{llm}.

In total, we generated $2,040$ solutions with {\sc Copilot} and $12,240$ with each of the eight other models, resulting in $210,120$ generated solutions overall.

\subsection{Validation}\label{sec:validation}
Each generated solution was tested to ensure its functional correctness following a three-step process.
At every step, if a solution was found to be invalid, it was excluded from subsequent stages of the experiment.
The process was as follows:

\textit{i) Local validation}. 
We filtered out code generations that included easy-to-spot errors, such as syntax errors or runtime errors, by using the small inputs we fetched earlier (see \citesec{tasks} - \emph{Test cases}). 
While this step was not strictly necessary, it quickly reduced the number of solutions to be validated in the next step;

\textit{ii) Leetcode validation}.
Next, we submitted the solutions to the Leetcode judge system using the Leetcode {\sc GraphQL\,API}, where they underwent a rigorous test suite managed by Leetcode. 
We kept the solutions that passed all the test cases or exceeded Leetcode's allocated time limit. 
The latter was kept to ensure that correct solutions that were too slow remained included in the benchmark;

\textit{iii) Exclusion of timeouts and other errors}.
Finally, we excluded the solutions that reported errors when executed with our benchmarking setup using the large inputs fetched beforehand.
We invalidated the code generations that took more than 10 seconds to run or raised an error. 
Most of the errors raised in this step were recursion errors caused by differences between Leetcode's Python interpreter and ours. 
Indeed, Leetcode's seemed to have a higher recursion limit than ours, which we set to $10,000$ instead of the default $1,000$ (we could not manage to set it any higher). 
Additional errors occurred because we developed our helper classes differently from Leetcode's implementation.
Although these classes are provided by Leetcode during the submission process, they are not publicly available.
Out of the $4,930$ invalidated solutions that were caught in this step, $4,863$ (98.6\%) were due to timeouts, $20$ (0.04\%) to recursion errors, and $47$ (0.1\%) to other errors.
Following this validation process, the initial set of $210,120$ generated solutions was pruned down to $7,481$ (3.6\%) valid solutions remaining across the $18$ \glspl{llm}. 

\subsection{Measuring run time}\label{sec:method_measure}
We measured the performance as the run time of the generated solutions using {\sf pytest-benchmark},\footnote{https://github.com/ionelmc/pytest-benchmark} which runs {\sf pytest} unit tests multiple times to obtain run times statistics.
The measurements were performed using parameters that ensured each solution ran at least $10$ times and for at least $1$ second in total.
We did not perform warm-up runs of the benchmarks, as we did not notice any significant difference in the measured time during preliminary testing. 
To facilitate the measurement protocol, the generated solutions were sorted into ``runs", based on the specific \gls{llm} and temperature that were used during the code generation.
Within each run, which was defined by a unique combination of model and temperature, the solutions were measured in sequence during a single program execution.
Furthermore, in every run, we added the canonical solutions we previously collected, which would run alongside the generated solutions.
This approach ensured that the same canonical solutions were executed in every run, allowing us to maintain measurement stability.
Specifically, we calculated the standard deviation of the canonical solution run times across all runs, thus providing a reliable measure of the variability of the measurement protocol.
We observed that over 96\% ($196$ out of $204$) of the canonical solutions had a standard deviation lower than $1/10^{th}$ of their average run time, which we deemed to be an acceptable level of variation.

The cluster we used to run the benchmark was the \mm{chiclet} cluster of the Grid5000 testbed.\footnote{\url{http://grid5000.fr}}
It hosts 2 AMD EPYC 7301, with 16 cores per CPU and 128GB of memory.
When using the node, all the cores of both CPUs were reserved, but only one was used at a time to maximize the stability of the measurement protocol.

\subsection{Replication package}
All the artifacts of this study, including our results, code, and datasets, are available in the following public repository: \url{https://zenodo.org/doi/10.5281/zenodo.7898304}.

\section{Data Analysis}\label{sec:data_analysis}
In this section, we describe the methods we adopted to analyze our results.
These methods fall into one of the two following categories: functional correctness and code performance.

\subsection{Functional Correctness}
The functional correctness of an \gls{llm} defines how much the \gls{llm} outputs code conforming to the program contract (as specified by the input prompt).
To evaluate the functional correctness of our \glspl{llm}, we computed their {\sf pass@k} metrics with $k=1$ and $k=10$, using the unbiased estimator proposed by Chen~\etal~\cite{chenEvaluatingLargeLanguage2021}.
The {\sf pass@k} unbiased estimator which, from {\sf k} samples produced, considers the test as successful if one of these samples passes all the tests, is computed as follows (with $n$ the total number of samples, $c$ the number of correct samples and $\mathbb{E}$ the expected value):
\vspace{-1mm}
\begin{equation}
pass@k := \underset{Problems}{\mathbb{E}}\left[1- \frac{\binom{n-c}{k}}{\binom{n}{k}}\right]
\end{equation}
\vspace{-1mm}

As Chen~\etal~\cite{chenEvaluatingLargeLanguage2021} suggest, we calculated the {\sf pass@k} for each temperature when evaluating an \gls{llm}'s functional correctness and considered the best one as the {\sf pass@k} for that \gls{llm}.

\subsection{Code Performance}
To measure the code performance, we considered three different metrics.
First, we used the memory usage reported by Leetcode. Then, we computed the median of the run times measured by {\sf pytest-benchmark} for every generated solution.
Lastly, to compare the \glspl{llm} solutions to human-submitted solutions, we also used the rank reported by Leetcode when validating the solution. 
This rank is a number between 0 and 100 that indicates the share of submitted solutions that are slower than the current solution (\eg if a solution has a rank of 90, it is faster than 90\% of the submitted solutions on Leetcode). 

To assess the \glspl{llm}' performances, we conducted pairwise comparisons as follows: For each pair of \glspl{llm}, we identified problems where both models generated more than 5 valid solutions. 
For each identified problem, we conducted a Student $t$-test on the mean run time of the generations to determine if there was a significant difference. 
Then, for each pair A-B of \glspl{llm}, we computed the ratio of problems where A's code was significantly faster than B's and where B's code was significantly faster than A's.

%% file: models.tex
\begin{tabular}{llcc}
\hline
\bf \gls{llm} Model & \bf Model family & \bf Size & \bf RQ1 \\ \midrule
GitHub Copilot & Codex & 11 & \checkmark\\
CodeGen-Mono 6B & CodeGen & 6 & \checkmark\\
CodeGen-Mono 2B & CodeGen & 2 & \checkmark\\
CodeGen-Mono 350M & CodeGen & 0.35 & \checkmark\\
CodeGen2.5-7B-mono & CodeGen2.5 & 7 & \\
CodeGen2.5-7B-instruct & CodeGen2.5 & 7 & \\
CodeLlama-7B-instruct & CodeLlama & 7 & \\
CodeLlama-7B & CodeLlama & 7 & \\
CodeLlama-7B-python & CodeLlama & 7 & \\
CodeLlama-13B-instruct & CodeLlama & 13 & \\
CodeLlama-13B-python & CodeLlama & 13 & \\
replit-code-v1-3b  & replit-code & 3 & \\
WizardCoder-pythin  & WizardCoder & 7 & \\
SantaCoder & Santacoder & 1.1 & \checkmark\\
StarCoder & StarCoder & 15.5 & \\
InCoder 6B & Incoder & 6 & \checkmark\\
InCoder 1B & Incoder & 1 & \checkmark\\
CodeParrot & Codeparrot & 1.5 & \checkmark\\
\end{tabular}

%% file: Results.tex
\section{Results}
\label{sec:results}
In this section, we summarize the key observations from our experiment and answer our research questions. 
In \autoref{table:pass_at_k_new_dataset}, you can also find the functional validity results of the different \glspl{llm} on both of our Leetcode datasets.
Our results are also available in the form of a companion notebook in our replication package. 
The companion notebook offers more insight into the results and additional graphs.

\begin{table}[ht] \centering
    \renewcommand{\arraystretch}{0.85}
    \begin{small}
        \input{pass_at_k_new_dataset.tex}
    \end{small}
    \vspace{2mm}
    \caption{Functional validity of the \glspl{llm} on Leetcode (by decreasing {\sf pass@1}, higher is better)}
    \label{table:pass_at_k_new_dataset}
\end{table}

\subsection{RQ1: Can Leetcode be used as a dataset and a benchmark platform for evaluating \glspl{llm}?}

\subsubsection{Can Leetcode problems be adopted as a dataset for \gls{llm} generation?}

\begin{figure}[t!]
    \centerline{\includegraphics[width=0.9\linewidth]{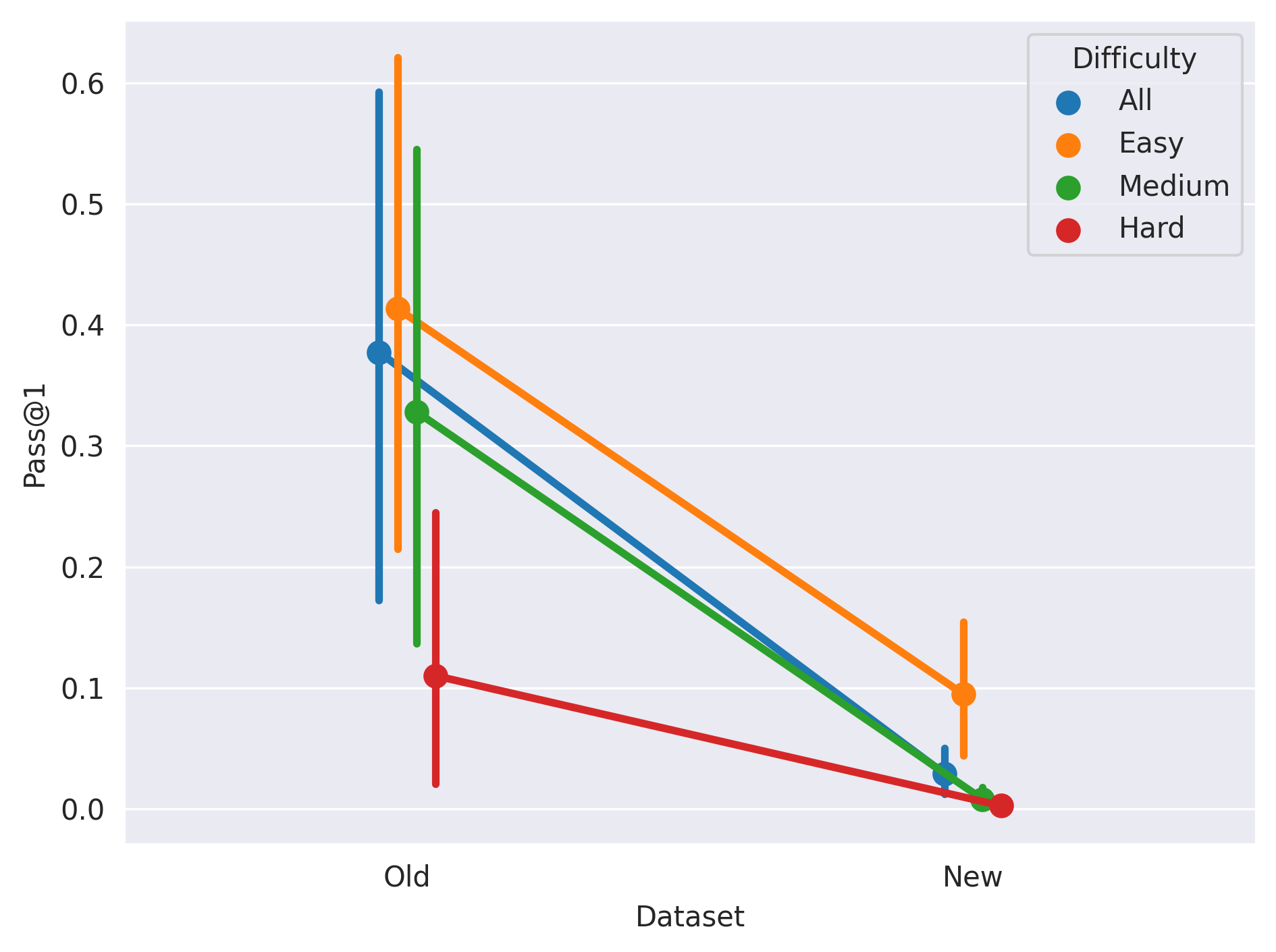}}
    \caption{Average {\sf pass@1} of the evaluated \glspl{llm} for every difficulty and dataset, with 95\% confidence interval (higher is better)}
    \label{fig:pass_at_1_by_difficulty_and_dataset}
\end{figure}

As on can observe in \autoref{fig:pass_at_1_by_difficulty_and_dataset}, the generated codes exhibit on a significant drop in functional correctness between the two datasets. 
The difference here is pretty staggering for every tested \gls{llm}, reporting on a tenfold decrease in {\sf pass@k}.
We believe this issue may stem from data contamination in the old dataset. 
Data contamination occurs when an \gls{llm} is assessed on data that was included in the training dataset, introducing bias into the evaluation process.
In our case, a significant number of questions in the old dataset are widely known and have been extensively shared on GitHub. 
For instance, a search for the prompt of the "3sum" Leetcode problem on GitHub yields approximately $4.000$ matches in public repositories.
These questions are also old enough to likely be included in the training datasets of the \glspl{llm} under study, as the majority of their training datasets have a cut-off date between 2021 and 2022. 
Due to this data contamination, \glspl{llm} tend to recite, reproducing verbatim source code when generating solutions.
This phenomenon is more pronounced when the prompt is highly specific and lacks contextual information, as seen in Leetcode prompts that closely match GitHub repositories.
The observed shift in functional validity between the two datasets could also arise from a genuine difference in the difficulty of the questions within each dataset. 
However, quantifying this last hypothesis proves to be challenging.

\subsubsection{Are Leetcode measurements reliable?}

\begin{figure}[t!]
    \centerline{\includegraphics[width=\linewidth]{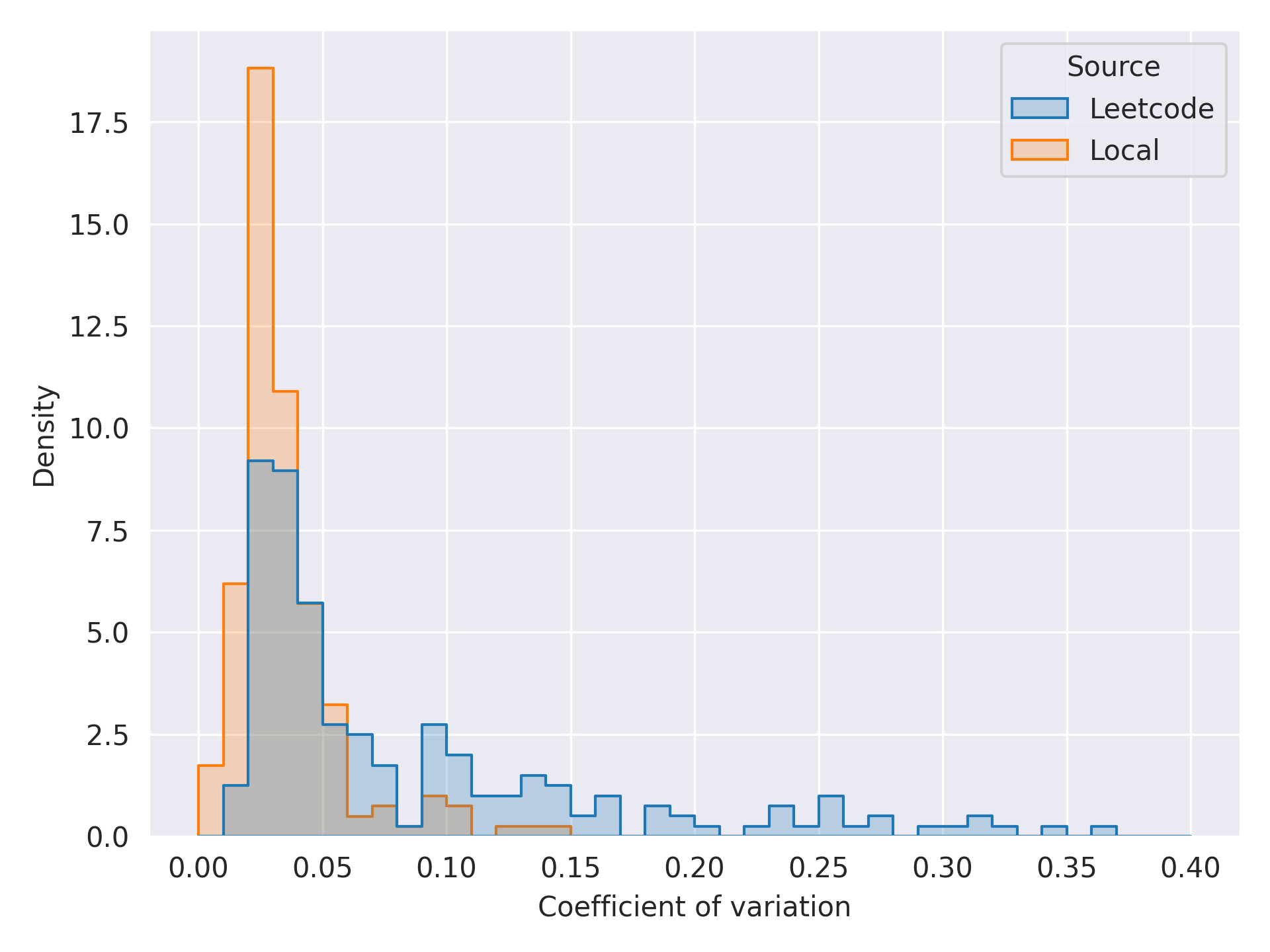}}
    \caption{Coefficicent of variation of the time measured by Leetcode and locally for every problem using canonical solutions}
    \label{fig:leetcode_run_time}
\end{figure}

\textbf{Run time.}
As reported in \autoref{fig:leetcode_run_time}, the coefficient of variation of the Leetcode measures ($0.089$) is slightly higher than the coefficient of variation of the local measures ($0.035$). 
This suggests that Leetcode's measuring setup is less suited to ensure accurate benchmarks.

\begin{figure}[t!]
    \centerline{\includegraphics[width=0.9\linewidth]{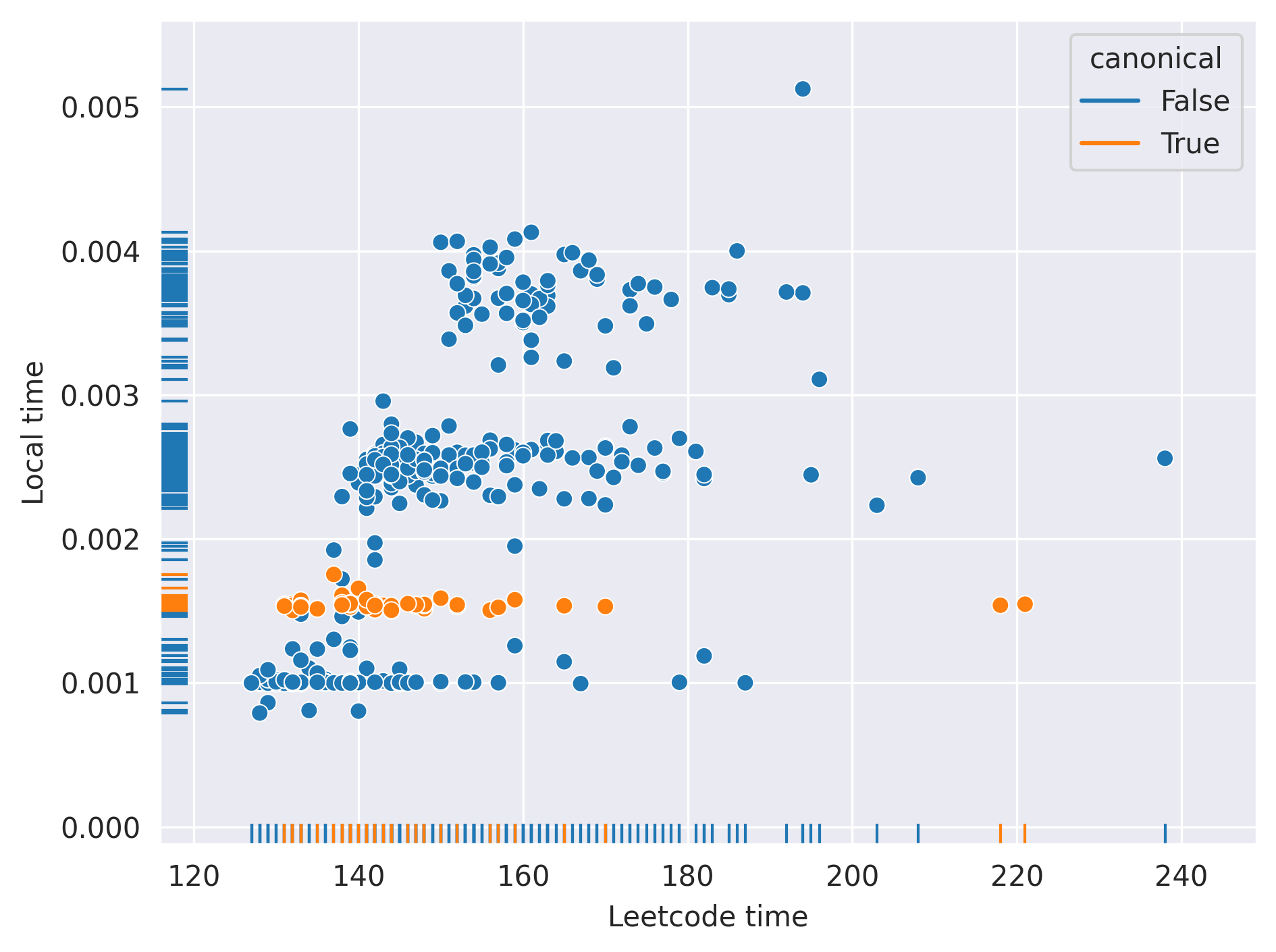}}
    \caption{Scatter plot of the measures done by Leetcode and locally for every generation for the problem "\textit{Difference between element sum and digit sum of an array}". Orange points are from multiple measures of the same canonical solution and serve as visual references for the measurement error}
    \label{fig:leetcode_vs_local_time_for_problem_difference}
\end{figure}

We also study the correlation between our local measurements and Leetcode's, and we notice two issues: (1) the times measured locally and by Leetcode only slightly correlate on average ($0.28$). 
For some problems, the measures are highly correlated ($> 0.8$) while, for others, they are almost not ($<0.2$).
This is more apparent when we look at the scatter plot showing the measures of some problems in \autoref{fig:leetcode_vs_local_time_for_problem_difference}. 
In this problem, there are four clusters of generations with a different locally measured time, but the clusters are indiscernible in terms of Leetcode time. 
This could be due to two main reasons. 
Firstly, as previously discussed, the variance of the measures from Leetcode is higher and as such, there is much more noise in the measures, decreasing the precision. 
Secondly, the tests we employed may be more focused on performance testing than Leetcode's. 
Notably, our test suite comprises only three tests featuring significantly large inputs, potentially accounting for certain disparities in the results.

Although still usable, relying on the time reported by Leetcode introduces some limitations due to its higher variance. 
Consequently, Leetcode's measures cannot allow us to discern the differences in run time between different code implementations as precisely as locally measured time.

\textbf{Memory usage.}
While we did not measure the memory ourselves, we observed the variation of the memory usage measure provided by Leetcode. We notice that the memory usage for the same solutions decreases over time. There is indeed a slight correlation of -0.24 between the day of the year we tested our solution and the memory usage.
The fact that the memory usage measure evolves renders comparisons between LLMs harder. While we could theoretically offset the memory usage when we detect changes over time, it would require testing the canonical solutions alongside the generated one on Leetcode.

\begin{figure}[t!]
    \centerline{\includegraphics[width=0.8\linewidth]{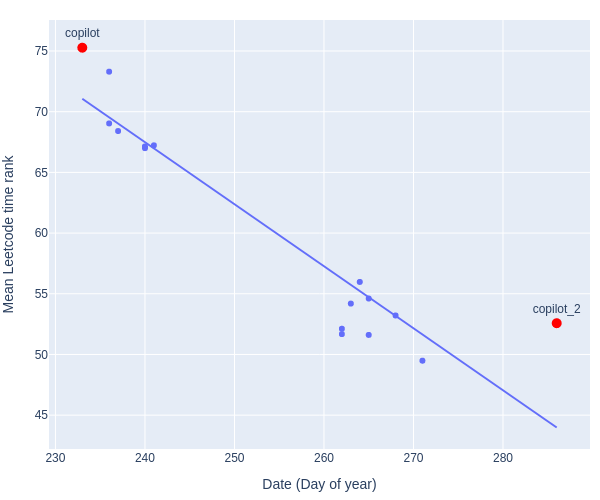}}
    \caption{Scatter plot of \gls{llm}'s rank and date they were tested on Leetcode. The two models in red are the same model tested on different dates}
    \label{fig:leetcode_rank}
\end{figure}

\textbf{Leetcode rank.}
The time ranking that Leetcode returns when we test a solution represents the share of submitted and valid solutions that are slower than ours. 
While this could be a great tool to rank \glspl{llm} among human-submitted solutions, we find that the ranking is heavily affected by our submissions and time. 
As you can see in \autoref{fig:leetcode_rank}, the overall rank of the \glspl{llm} we tested decreases over time. 
To verify this, we tested {\sc GitHub Copilot} twice, once as the first \gls{llm}, and a second time after testing all the \glspl{llm}.
The first test of {\sc Copilot} has an overall rank of 77, and the second test ranks down to 54, despite it being tested with the same solutions.
This effect of the rank evolving becomes obvious when you consider that the rank is determined using all the previously accepted solutions, including ours.
This means that by testing thousands of our solutions on Leetcode, we are actively changing the ranks of future tests.

\begin{formal}
\textbf{RQ1}: The evaluation of \glspl{llm} using Leetcode questions as a dataset presents some challenges.
Although Leetcode's questions could serve as a valuable dataset akin to {\sc HumanEval}, limitations arise due to the constraint that only the problems published after the \gls{llm}'s training dataset formation are usable for evaluating the \gls{llm} in question.
This creates potential difficulties in reproducibility, particularly as new \glspl{llm} emerge, especially if they do not exclude Leetcode problems from their training datasets~\cite{jacoviStopUploadingTest2023}. 
Additionally, while Leetcode's provided metrics, such as run time, memory usage, and rank may offer practicality in various scenarios, their usability and reliability are questioned when compared to more traditional measurement methods. 
The presence of these challenges emphasizes the need for careful consideration and scrutiny when adopting Leetcode to evaluate \glspl{llm}.
\end{formal}

\subsection{RQ2: Are there notable differences in performances between \glspl{llm}?}

\begin{figure}[t!]
\centerline{\includegraphics[width=0.9\linewidth]{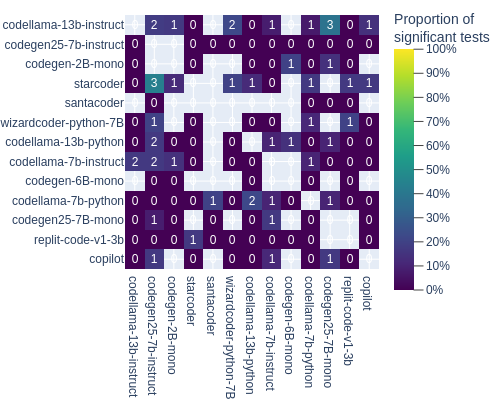}}
\caption{Number of problems where an \gls{llm} (row) is better than another (column)}
\label{fig:pairwise_comparison}
\end{figure}

The pairwise comparison depicted in \autoref{fig:pairwise_comparison} reveals subtle distinctions in the performances of various \glspl{llm}.
Notably, some models, such as StarCoder and the CodeLlama model with 13B parameters specialized in Python, consistently exhibit slightly superior results compared to others. 
Despite these observed variations, the mean Cohen's $d$ effect size measures a mere $0.024$, a statistically insignificant magnitude. 
This suggests that the practical impact of these differences on the mean speed of code generation is remarkably small.
For instance, when comparing CodeLLama-13-instruct and CodeGen25-7B-mono, CodeLLama outperforms the latter in a statistically significant manner in 3 problems out of 8. However, it is crucial to note that the mean performance difference between these models is a mere $0.02$ standard deviation.
It thus seems that improving an \gls{llm} in terms of functional validity does not significantly impact the performance of the code it generates. 
This may be due to different factors, such as the fact that most \glspl{llm} share the same datasets or that they are trained to produce valid code and not fast code.
Improving the performance of an \gls{llm} could be done by curating a training dataset of only efficient code and fine-tuning one of the foundational models we used, or by using reinforcement learning to "teach" the model to produce better code. 
Madaan~\etal produced an \gls{llm} that could improve the performance of code~\cite{madaanLearningPerformanceImprovingCode2023}. 
This \gls{llm} could be leveraged in a generation pipeline to directly improve the generated code. %

\begin{samepage}
\begin{formal}
\textbf{RQ2}: Our analysis uncovers statistical differences in the performance of generated code among different \glspl{llm}.
However, the effect size, as measured by Cohen's $d$, is so negligible that it raises questions about the practical significance of these differences. 
Despite some models consistently outperforming others, the overall impact on the mean efficiency of \gls{llm}-generated code appears to be minimal.
\end{formal}
\end{samepage}

\subsection{RQ3: Is there an effect of the functional validity of the \gls{llm} and its temperature on the generated code's performance?}

\textbf{Functional validity.} 
When calculating for every problem the correlation between the success rate of the \gls{llm} that generated the solution and the run time of the solution, we find that there is only a very slight negative correlation ($-0.08$) between the success rate and the performance.
There is close to no correlation ($-0.11$) observed between the success rate of the model and the variation in the performance of the generated code.

\begin{figure}[t!]
\centerline{\includegraphics[width=\linewidth]{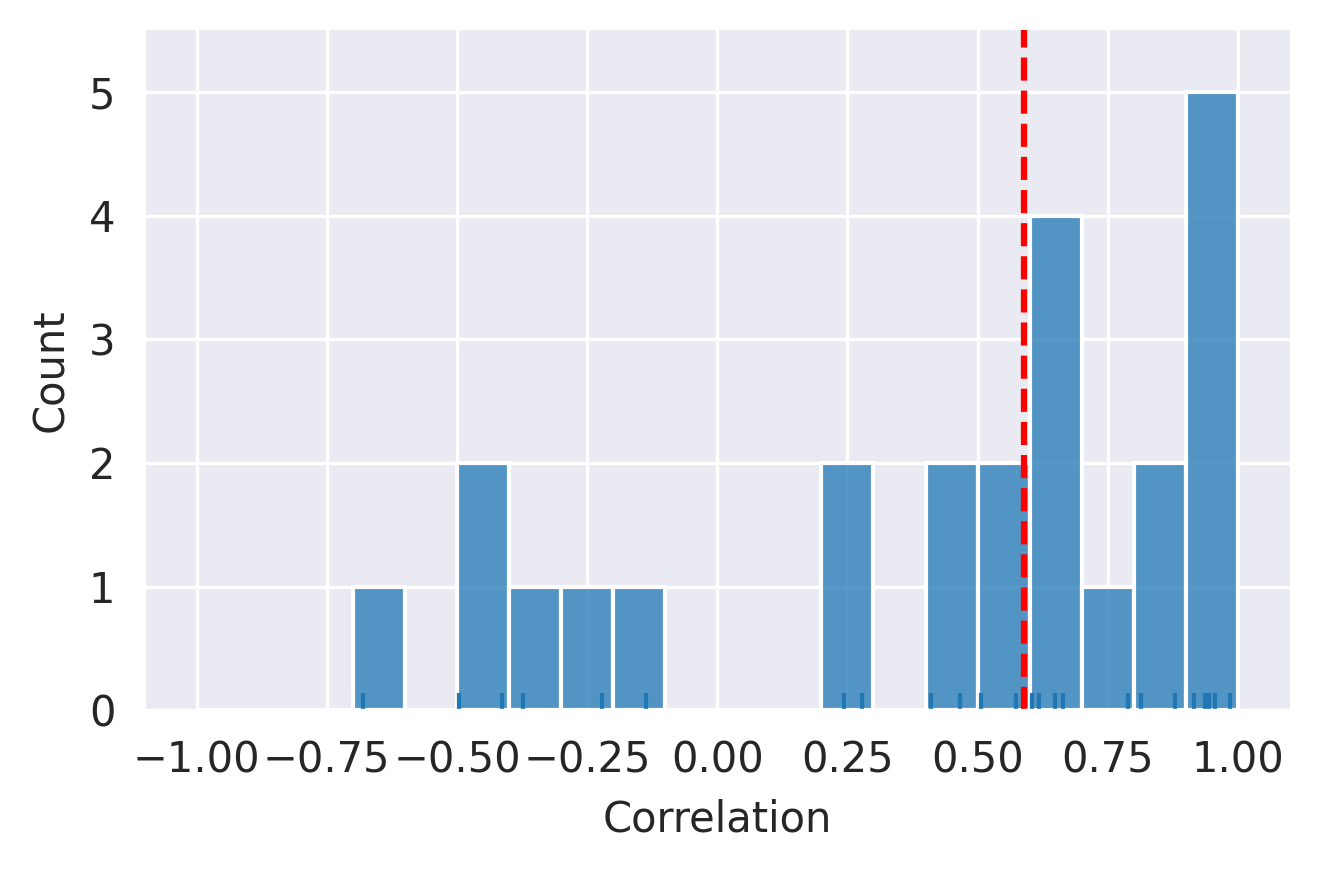}}
\caption{Distribution of correlations for every problem between the temperature and the variation of the performance. The red line is the median}
\label{fig:corr_temp_run_time_std}
\end{figure}

\textbf{Temperature.} 
There is no correlation ($0.05$) observed between the temperature of the generations and the performance of the generated code. 
This means that the temperature does not affect how fast the solutions are.
However, we observe that temperature is moderately correlated ($0.41$) with higher variations in performances. 
This means that higher temperatures tend to increase the variation in performance across generations. 
The complete distribution of the correlation for each of the 24 problems can be seen in \autoref{fig:corr_temp_run_time_std}. %
So, while increasing the temperature leads to a lower success rate~\cite{doderleinPilotingCopilotCodex2022}, it can help find a faster solution with an extended exploration of generations.
The fact that the temperature increases the variation in performances comforts the idea that higher temperatures lead to more diverse outcomes.

\begin{samepage}
\begin{formal}
\textbf{RQ3}: Our analysis of \glspl{llm} indicates that the quality of generated code does not have a substantial impact on its performance. 
However, we observed that modifying temperature settings within an \gls{llm} significantly affects the diversity of code performances produced. 
This implies that, while code validity may not be a decisive factor in performance, adjusting temperature settings can be a valuable strategy to enhance the variety of outcomes in code generation processes.
\end{formal}
\end{samepage}

\subsection{RQ4: How fast are LLMs compared to humans ?}
As previously stated, the Leetcode time ranking evolves, so we chose to compare the second model we tested on Leetcode with humans ({\sc Copilot} being the first, it did not have enough generations overall because of its lack of a temperature setting). 
The results of this comparison are depicted in \autoref{fig:leet_rank_early_model}.
The comparison is done using the Leetcode ranking, with the assumption that most of the previous submissions were made by humans.

We observe in \autoref{fig:leet_rank_early_model} that the solutions generated from \glspl{llm} are faster than most previous submissions with a mean rank of 73\%, and that it even generated some solutions that were faster than 95\% of the previous submissions.

\begin{samepage}
\begin{formal}
\textbf{RQ4}: It seems that the \glspl{llm} are faster than most of the human solutions on Leetcode, on average. 
If the \gls{llm} we tested were in an actual competition, his valid solutions would be on average faster than 73\% of the other solutions on Leetcode.
\end{formal}
\end{samepage}

%% file: pass_at_k_new_dataset.tex
\begin{tabular}{lrr}
\toprule
\bf \gls{llm} Model & \bf Pass@1 & \bf Pass@10 \\
\midrule
StarCoder & 0.095 & 0.132 \\
CodeLlama-13B-python & 0.093 & 0.201 \\
GitHub Copilot & 0.092 & 0.196 \\
CodeLlama-7B-instruct & 0.082 & 0.191 \\
CodeLlama-13B-instruct & 0.078 & 0.206 \\
WizardCoder-python-7B & 0.075 & 0.157 \\
CodeGen2.5-7B-mono & 0.066 & 0.147 \\
CodeGen2.5-7B-instruct & 0.062 & 0.142 \\
CodeLlama-7B-python & 0.047 & 0.172 \\
CodeGen-6B-mono & 0.045 & 0.113 \\
CodeGen-2B-mono & 0.038 & 0.103 \\
replit-code-v1-3b & 0.025 & 0.083 \\
InCoder-6B & 0.021 & 0.064 \\
SantaCoder & 0.015 & 0.064 \\
CodeLlama-7B & 0.014 & 0.015 \\
InCoder-1B & 0.012 & 0.039 \\
CodeGen-350M-mono & 0.007 & 0.039 \\
CodeParrot & 0.002 & 0.015 \\
\bottomrule
\end{tabular}

%% file: Discussion.tex
\section{Discussion}\label{sec:discussion}
In this section, we discuss the results reported in the previous section, their implications, and the limitations of our study.

\subsection{Discussion of the results}
\textbf{On the Leetcode measures and usability.}
The data contamination issue we unveiled poses a significant challenge in the evaluation of \glspl{llm}, as it prevents an accurate evaluation of their real performances. 
Because Leetcode's problems are not filtered from the training datasets, as research on \glspl{llm} continues, even the newer problems might contaminate future training datasets, thus rendering reproduction of our study harder~\cite{jacoviStopUploadingTest2023}. 
This conclusion also holds for any study using Leetcode as an evaluation dataset, such as~\cite{bubeckSparksArtificialGeneral2023, doderleinPilotingCopilotCodex2022, nguyenEmpiricalEvaluationGitHub2022}. 
However, we believe that the methodologies employed and the conclusions drawn would likely hold validity with alternative sets of questions from Leetcode or other performance-oriented datasets. 
This suggests that future studies seeking to replicate our findings would primarily need to change the dataset employed for assessing \glspl{llm}.
Addressing the data contamination concern could involve leveraging pre-filtered evaluation datasets, such as {\sc HumanEval}, already separated from the training processes, and repurposing them into performance evaluation datasets.

\begin{figure}[t!]
\centerline{\includegraphics[width=\linewidth]{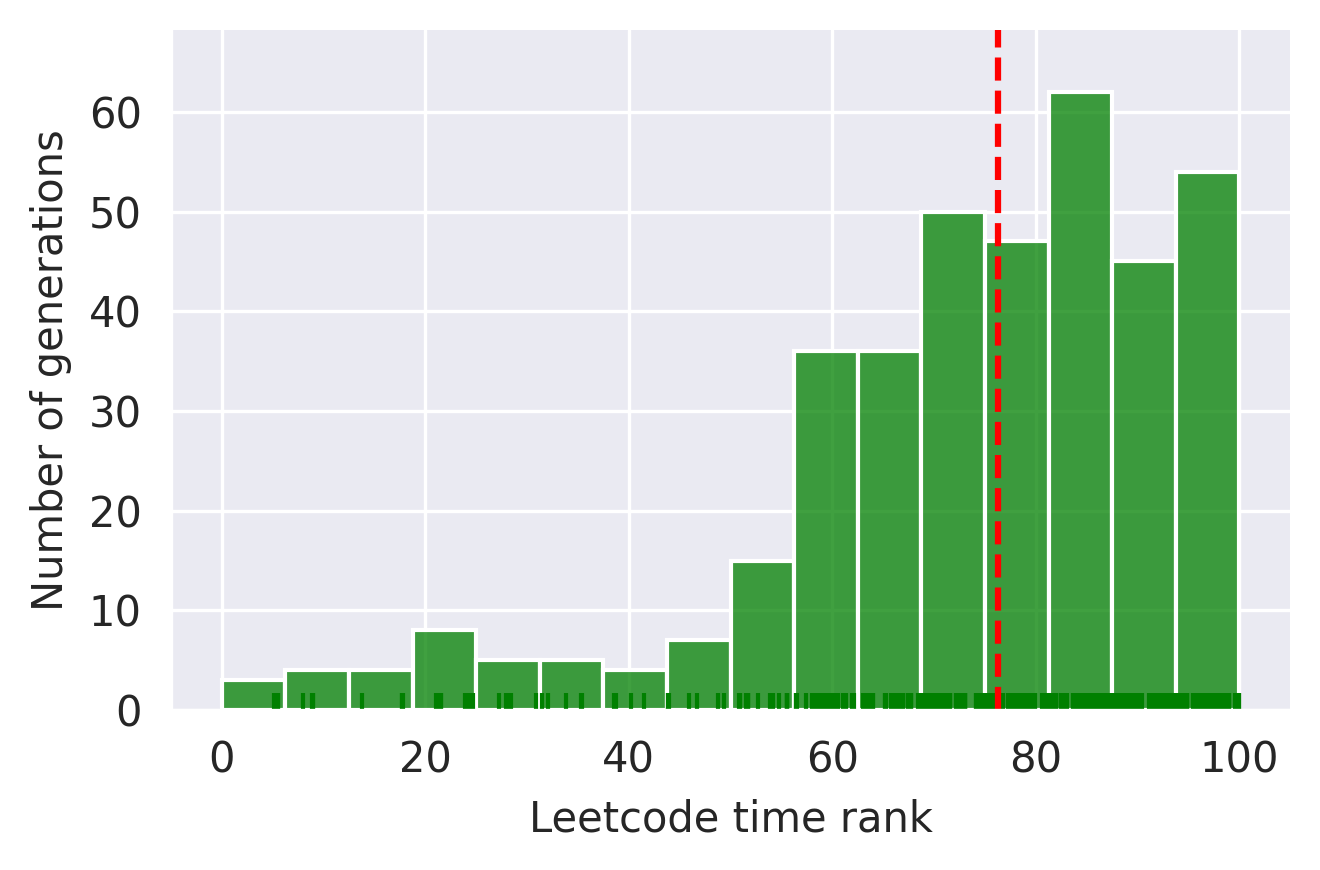}}
\caption{Distribution of the ranking for the CodeGen-6B-mono model}
\label{fig:leet_rank_early_model}
\end{figure}

\textbf{On functional correctness}.
The ranking of the functional correctness of the \glspl{llm} is consistent with the previous evaluations of the \glspl{llm} on {\sc HumanEval}~\cite{chenEvaluatingLargeLanguage2021, CodeparrotCodeparrotHugging2022, nijkampCodeGenOpenLarge2022, allalSantaCoderDonReach2023,li2023starcoder, luoWizardCoderEmpoweringCode2023, roziereCodeLlamaOpen2023}. 
However, it seems that InCoder performs worse than presented in its introductory paper~\cite{friedInCoderGenerativeModel2022}, which may be due to our experimental protocol---using it only for left-to-right generation instead of infilling like it was built for.  %

We observe that Leetcode's problems seem harder for the \glspl{llm} to solve than {\sc HumanEval}'s problems. 
Indeed, StarCoder, the model that performed the best with a {\sf pass@1} of $0.09$, had a {\sf pass@1} of $0.408$ on {\sc HumanEval}~\cite{li2023starcoder}. 
We believe this is due to multiple factors. 
First, our Leetcode prompts and expected solutions are longer than in {\sc HumanEval} (the average length of solution in {\sc HumanEval} is 180 characters \emph{vs} 425 characters in our dataset), thus increasing the chance of a generation to fail. 
Indeed, having to generate more code can lead to a higher chance of making a mistake, as shown by a correlation of $-0.30$ between the solution's length and the success rate of the problem.
Second, Leetcode problems come from programming competitions and need a lot of thinking to be solved. 
The causal generation of the \glspl{llm} does not allow a "thinking" process to happen, which for harder problems causes a drop in functional validity. 
This could be solved by making the \glspl{llm} mimic human thinking with methods, such as Chain-of-Thought prompting~\cite{weiChainofThoughtPromptingElicits2023}.

\textbf{On the performance of generated code}. 
To the best of our knowledge, our methodology for evaluating \glspl{llm} based on the performance of generated code is novel and could serve as a benchmarking approach for future studies involving new \glspl{llm} and datasets.
While we aimed for a singular performance score for \glspl{llm}, similar to pass@k, the varying rates at which \glspl{llm} generate valid solutions posed a challenge, leading us to employ pairwise comparisons.
An improvement to our method could involve using an optimal solution as a reference point and comparing the speed of each generated solution to the optimal solution's speed. 
Speed would thus be expressed as a factor relative to the optimal time, addressing the issue of problems having different time scales and potentially laying the groundwork for a more comprehensive "performance score."

The low success rate of \glspl{llm} on Leetcode posed a considerable challenge for their comparison. 
Among the 204 problems, only 24 had \emph{(i)} valid solutions from at least 10 different models (out of the 18 assessed \glspl{llm}) and \emph{(ii)} at least 10 valid solutions in total (see to the companion notebook). 
This low success rate complicated pairwise comparisons, as numerous models could not be compared due to an insufficient number of common problems with a substantial number of generated solutions.

Our discoveries offer valuable insights for developers in their choice of \glspl{llm}. 
For example, when developers are considering an LLM for tasks like code generation, such as with GitHub Copilot, the performance of the generated code may be an important concern.
With our findings, developers can be assured that there is no significant variance in the performance of code generated by different \glspl{llm}.
This means that if they aim to have fast code, 
there's no necessity to consistently opt for the largest model available; instead, a smaller one suffices.
We also hope that our findings will incentivize further research on building new \glspl{llm} that produce even more efficient code.

\subsection{Limits and Threats to Validity}
Regarding functional correctness, although we did our best to generate code in the most optimal conditions, some changes to the input prompt (\ie adding examples and constraints), or the configuration of the models may have changed the performance of the \glspl{llm}.
However, we believe that this should not significantly impact the validity of our experiment, as all the models were configured similarly.

{\sc GitHub Copilot} being a closed-source tool, it might be retrained without the community's knowledge, which could potentially lead to a modification in the tool's performance in upcoming experiments.

The majority of our validation and benchmarking process relied on Leetcode's test suite and online judge system. 
Thus, it is possible that some big test cases we extracted for the benchmarking favored some types of implementations over others. 
We mitigated this issue by systematically fetching three test cases and by having a large dataset of problems to generate from.
During our experiments, we also tried to generate plausible inputs using random generators, but this method did not yield satisfying results because randomly generating data structures with specific shapes, or properties (\eg generating valid regular expressions) is difficult to achieve. 

One gap in our study is that we do not consider memory usage at all when studying the \glspl{llm}. 
This is because our benchmarking setup did not allow for memory monitoring and doing otherwise would have cost us a lot of time. 
Moreover, we believe our results obtained with only the run time to be self-sufficient.

The way we compared an \gls{llm} to a human using Leetcode rankings gives a good idea of where the \gls{llm} stands in terms of performance. 
However, because the ranking evolves and we have no information about the population the \gls{llm} is ranked against, the results here should be taken with a grain of salt.

It is also important to note that Leetcode as an evaluation dataset suffers from some issues. 
As we only evaluate the \glspl{llm} on algorithmic problems, the performances of the \glspl{llm} are hard to generalize across all programming fields. 
However, this is difficult to improve on for similar reasons to {\sc HumanEval}: we only have a limited context size and have to make the \gls{llm} generate in one go a completion to some code, which must be self-contained (meaning, all the information needed to generate the solution must be in the prompt). 
Also, in terms of performance, it is difficult to find another kind of self-contained code than algorithmic problems that have, such variations in performance. 
While studying the \glspl{llm} on SQL generation could be a good idea, it would not fit into our studies with generalist programming languages.

Regarding Leetcode as a platform, we are heavily dependent on it for validating the generated solutions and are limited by its daily rate limits of $1,000$ submissions per account. 
For future studies, it would be great to consider alternatives to Leetcode, such as the project CodeNet~\cite{puriCodeNetLargeScaleAI2021}, which does not have as many restrictions as Leetcode.

%% file: Related.tex
\section{Related Works}\label{sec:rw}
Previous research has investigated various aspects of \glspl{llm} for code-related tasks, including the security of their suggestions~\cite{pearceAsleepKeyboardAssessing2022, sandovalSecurityImplicationsLarge2022, perryUsersWriteMore2022}, the prevalence of bugs in the generated code~\cite{jesseLargeLanguageModels2023}, how developers interact with them ~\cite{perryUsersWriteMore2022, vaithilingamExpectationVsExperience2022, barkeGroundedCopilotHow2022} or just the quality and correctness of the code they generate~\cite{yetistirenAssessingQualityGitHub2022, doderleinPilotingCopilotCodex2022, nguyenEmpiricalEvaluationGitHub2022,  liu2024your}. 
There have also been efforts to measure the efficiency of \glspl{llm} through the creation of benchmarks for comparing them, such as {\sc HumanEval}\cite{chenEvaluatingLargeLanguage2021}, MBPP~\cite{austinProgramSynthesisLarge2021},  {\sc CoderEval}~\cite{yuCoderEvalBenchmarkPragmatic2023}, APPS~\cite{hendrycksMeasuringCodingChallenge2021}, {\sc CodeXGLUE}~\cite{luCodeXGLUEMachineLearning2021} or {\sc ReCode}~\cite{wangReCodeRobustnessEvaluation2022}. 
Xu~\etal~\cite{xuSystematicEvaluationLarge2022} also led a comparative evaluation of multiple \glspl{llm} for code including Codex and Codeparrot.

Leetcode, while being just a coding competition platform, is also used as a dataset to evaluate the capabilities of \glspl{llm} on programming tasks. 
Döderlein~\etal~\cite{doderleinPilotingCopilotCodex2022} measured the performances of {\sc Copilot} and Codex on Leetcode and the effects of changing the prompts. 
Nguyen and Nadi~\cite{nguyenEmpiricalEvaluationGitHub2022} studied {\sc Github Copilot}'s code suggestions on Leetcode problems and the complexities of its generated code. 
Vasconcelos~\etal~\cite{vasconcelosGenerationProbabilitiesAre2023} studied the effects of highlighting the uncertainty of AI-powered code completions using Leetcode problems and Codex.

Various other methods have also been employed to investigate the impact of temperature on the generated code, apart from the approach we proposed: Chen~\etal~\cite{chenEvaluatingLargeLanguage2021} evaluated the best temperature of Codex in terms of {\sf pass@k}. 
Austin~\etal~\cite{austinProgramSynthesisLarge2021} also studied the effects of the temperature on the performance of their model.
Christopoulou~\etal~\cite{christopoulouPanGuCoderProgramSynthesis2022} also studied the effects of the temperature and nucleus-sampling on their \gls{llm}.
Döderlein~\etal~\cite{doderleinPilotingCopilotCodex2022} highlight the importance of correctly tuning the temperature of a model when using it to generate code. 
The research led by Aghakhani~\etal~\cite{aghakhaniTrojanPuzzleCovertlyPoisoning2023} shows that poisoned models suggest insecure code more often as the temperature increases.
Our results demonstrate that increasing the temperature also increases the chance of generating slow and inefficient code.

On the subject of performance, Madaan~\etal~\cite{madaanLearningPerformanceImprovingCode2023} fine-tuned \glspl{llm} to make them improve the performance of code.
Multiple other techniques have been proposed for automatically improving the performance of code using \glspl{llm}~\cite{chenLearningImproveCode2022, gargDeepDevPERFDeepLearningbased2022, chenSupersonicLearningGenerate2023}. 
Our contribution is the first---as far as we know---to investigate the differences in the performance of \gls{llm}-generated code. 
Future \glspl{llm} that would be adapted with these performance-improving techniques could also be compared using our methodology.
Regarding recitation, few works have been done on this subject, but Yan~\etal~\cite{yanWhyGenExplainingMLpowered2022} proposed a method to detect cases of recitations in \glspl{llm} using inference fingerprinting. 
Jacovi~\etal~\cite{jacoviStopUploadingTest2023} explained why data contamination in \glspl{llm} was problematic and proposed methods to mitigate it.

To summarize, our paper evaluates the performance of code generated by various \glspl{llm} and investigates differences in the performance of the generated code across models on Leetcode problems, which, to the best of our knowledge, has not been done previously.

%% file: Conclusion.tex
\section{Conclusion}\label{sec:conclusion}
In this study, we presented a comprehensive analysis of the performance of code generated by various \glspl{llm} using a novel methodology that measures and compares the runtime speed of solutions to algorithmic problems. 
Our findings suggest that the performance of the generated code is largely similar across different models, regardless of their size or training data. 
Furthermore, increasing the temperature parameter during code generation leads to a greater variance in performance, though not necessarily to better or worse solutions on average.
We also critically evaluated the suitability of Leetcode as a dataset and benchmark platform for assessing \glspl{llm}. 
The results indicate that, while Leetcode's problems are suitable for performance evaluation, their measures should be used cautiously due to issues with stability and reliability. 
Additionally, we observed that the use of newer Leetcode problems is essential to avoid data contamination and ensure the validity of \gls{llm} evaluations.

This work opens up several avenues for future research, including the development of performance-oriented training datasets and the fine-tuning of \glspl{llm} for performance improvement. 
As the field of AI-assisted programming continues to evolve, studies such as ours will play a critical role in understanding and enhancing the capabilities of \glspl{llm}. %